\documentclass[11pt,twoside]{article}

\usepackage{asp2006}
\usepackage{epsf}
\usepackage{lscape}

\begin{document}
\title{KPD0005+5106:\\ Hottest DO White Dwarf Much Hotter Than Assumed}
\author{K. Werner,$^1$ T. Rauch,$^1$ and J. W. Kruk$^2$}
\affil{$^1$Institute for Astronomy and Astrophysics, Kepler Center for
  Astro and Particle Physics, University of T\"ubingen, Sand~1, 72076 T\"ubingen, Germany\\
$^2$Department of Physics and Astronomy, Johns Hopkins University, Baltimore, MD 21218, U.S.A.
}  

\begin{abstract}
KPD0005+5106 is the hottest known helium-rich white dwarf. We have identified
\ion{Ne}{viii} lines in UV and optical spectra and conclude that it is
significantly hotter than previously thought, namely $T_{\rm eff}=200\,000$~K
instead of 120\,000~K. This is a possible explanation for the observed
hard X-ray emission as being of photospheric origin. Concerning its
evolutionary state, we suggest that KPD0005+5106 is not a descendant of a
PG1159 star but more probably related to the O(He) stars and
RCrB stars.
\end{abstract}

\section{Discovery of Ne\,\bf{\sc viii} lines and new temperature determination}

The hottest helium-rich white dwarfs exhibit lines of ionized helium in
their optical spectra and they are classified as DO white
dwarfs. Currently forty DOs are known. The coolest one has $T_{\rm
eff}=40\,000$~K and the hottest one is KPD0005+5106.  From an analysis
of optical and \emph{Hubble Space Telescope} FOS spectra $T_{\rm
eff}=120\,000$~K and $\log g=7$ has been derived (Werner et al.\@ 1994)
for KPD0005+5106. The Sloan Digital Sky Survey has recently revealed
another DO with $T_{\rm eff}=120\,000$~K (H\"ugelmeyer et al.\@ 2006).

In the \emph{Far Ultraviolet Spectroscopic Explorer} (FUSE) spectra of
the hottest PG1159 stars ($T_{\rm eff}>140\,000$~K) we have recently
discovered absorption lines of \ion{Ne}{viii} (Werner et al.\@
2007). The same features were also discovered in KPD0005+5106
(Fig.\,1). Emission lines in the optical spectrum are also from
\ion{Ne}{viii} and not, as previously thought, from superionized (i.e.,
non-thermally excited) \ion{O}{viii}. Non-LTE line profile fits to the
\ion{Ne}{viii} lines of KPD0005+5106 yield $T_{\rm eff}=200\,000$~K and
$\log g=6.5$. A reassessment of the \ion{He}{ii} line spectrum confirms
these parameters (Werner et al.\@ 2007). Hence, KPD0005+5106 is by far
the hottest He-rich white dwarf. In fact, it is the hotter than any
other white dwarf or central star of planetary nebula. It is only
rivaled by the exotic PG1159 star H1504+65 (Werner et al.\@ 2004, and
these proceedings).

\section{Nature of the observed X-ray emission}

ROSAT has revealed two X-ray emission components of
KPD0005+5106. \emph{Chandra}  observations of the soft component
(20--80\,\AA) prove that it is of photospheric origin (Drake \& Werner 2005)
and not, as claimed earlier, coronal (Fleming et al.\@ 1993). Deposing
the idea of non-photospheric \ion{O}{viii} lines is in accordance with
the deposition of the corona.  The hard component (at 12\,\AA) remained
unexplained (O'Dwyer et al.\@ 2003). In the light of the newly
determined, extreme $T_{\rm eff}$ it should be investigated whether this
component is of photospheric origin, too.

\section{Evolutionary context}

KPD0005+5106 is located on a  0.7\,M$_\odot$ Wood \& Faulkner (1986)
post-AGB track, just before the ``knee'', at the hot end of the white
dwarf cooling sequence. This position is well within the PG1159-class
domain of the HR diagram, hence, it cannot simply be a PG1159
descendant. The immediate progenitor could have been a O(He) star. This
spectral class consists of five hot stars ($T_{\rm
eff}=100\,000-140\,000$~K, $\log g=5.5-6.5$, Rauch et al.\@ 1998, and
these proceedings) with almost pure He atmospheres. KPD0005+5106 as well
as the O(He) stars might be descendants of RCrB stars.

\begin{figure}[!t]
\begin{center}
\epsfxsize=\textwidth  \epsffile{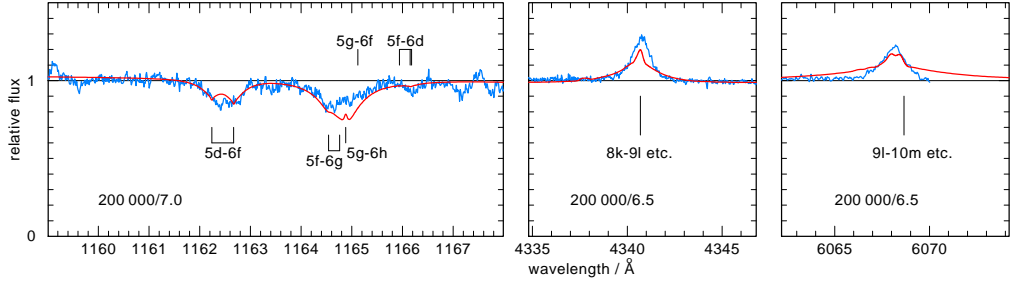}
\caption{Identification of \ion{Ne}{viii} lines in KPD0005+5106. 
Overplotted are computed profiles from models with $T_{\rm eff}$ and
$\log g$ as depicted.  \emph{Left panel:} Several lines of the $n=5
\rightarrow 6$ transition are detected in the FUSE
spectrum. \emph{Middle and right panels:} Optical spectral regions where
we identified the \ion{Ne}{viii} $n=8 \rightarrow 9$ and $n=9
\rightarrow 10$ transitions.}
\end{center}
\end{figure}

\acknowledgements T.R.\@ is supported
by the \emph{German Astrophysical Virtual Observatory} project of the
German Federal Ministry of Education and Research under grant
05\,AC6VTB. J.W.K.\@ is supported by the FUSE project, funded by NASA
contract NAS5-32985.

\end{document}